# Picoflare jets power the solar wind emerging from a coronal hole on the Sun

L. P. Chitta[1*], A. N. Zhukov[2,3], D. Berghmans[2], H. Peter[1], S. Parenti[4], S. Mandal[1], R. Aznar Cuadrado[1], U. Schühle[1], L. Teriaca[1], F. Auchère[4], K. Barczynski[5,6], É. Buchlin[4], L. Harra[5,6], E. Kraaikamp[2], D. M. Long[7,8], L. Rodriguez[2], C. Schwanitz[5,6], P. J. Smith[7], C. Verbeeck[2], D. B. Seaton[9]

[1]Max-Planck-Institut für Sonnensystemforschung, 37077 Göttingen, Germany

[2]Solar-Terrestrial Centre of Excellence, Solar Influences Data analysis Centre, Royal Observatory of Belgium, 1180 Brussels, Belgium

[3]Skobeltsyn Institute of Nuclear Physics, Moscow State University, 119991 Moscow, Russia

[4]Institut d'Astrophysique Spatiale, Centre National de la Recherche Scientifique, Université Paris-Saclay, 91405, Orsay, France

[5]Physikalisch-Meteorologisches Observatorium Davos, World Radiation Center, 7260 Davos Dorf, Switzerland

[6]Eidgenössische Technische Hochschule Zürich, 8093 Zürich, Switzerland

[7]Mullard Space Science Laboratory, University College London, Dorking, Surrey RH5 6NT, UK

[8]Astrophysics Research Centre, School of Mathematics and Physics, Queen's University Belfast, Belfast, BT7 1NN, Northern Ireland, UK

[9]Southwest Research Institute, Boulder, CO 80302, USA

*Corresponding author; E-mail: chitta@mps.mpg.de.

**Coronal holes are areas on the Sun with open magnetic field lines. They are a source region of the solar wind, but how the wind emerges from coronal holes is not known. We observed a coronal hole using the Extreme Ultraviolet Imager on the Solar Orbiter spacecraft. We identified jets on scales of a few 100 kilometers, which last 20 to 100 seconds and reach speeds of ~100 kilometers per second. The jets are powered by magnetic reconnection and have kinetic energy in the picoflare range. They are intermittent but widespread within the observed coronal hole. We suggest that such picoflare jets could produce enough high-temperature plasma to sustain the solar wind and that the wind emerges from coronal holes as a highly intermittent outflow at small scales.**

Hot plasma continuously escapes the Sun into the heliosphere, forming the solar wind (*1, 2*). Fast solar wind streams (those with speeds greater than 500 km s$^{-1}$) have been traced to large (mostly polar) coronal holes–features formed by open magnetic fields on the Sun that appear dark in coronal images taken at extreme ultraviolet (EUV) or X-ray wavelengths (*3–5*). Small coronal holes that form adjacent to active regions (areas on the Sun with strong magnetic fields including sunspots) could be the source regions of slower wind, with speeds less than 500 km s$^{-1}$ (*6, 7*). The physical origin and acceleration mechanism of solar wind from coronal holes are debated (*5, 8, 9*). Potential drivers include wave dissipation and turbulent cascades (*10–12*), or interchange reconnection between open and closed magnetic field lines at the coronal base (*13–15*). Jets around 0.1 MK emerging from the transition region network at the base of the corona, common in coronal hole regions, could sustain the solar wind (*16*). Coronal counterparts to the network jets are rare (*17, 18*), so it is unclear whether the cooler network jets can efficiently supply hot plasma to drive the solar wind.

Previous EUV imaging has shown signatures of outflows (jets) from plumes (bright ray-like features embedded in coronal holes) and the ambient, inter-plume, coronal hole regions (*19, 20*). EUV images with spatial resolution of about 1000 km, have shown that plumes have sub-structures consisting of jets with widths as small as 1500 km (*21–23*). These jets appear to be universal in the corona: they emerge not only from coronal hole plumes but also from quiet regions and active regions of the Sun's atmosphere (*24*). Jets from plumes could sustain the solar wind mass flux during the minimum of the 11-year solar activity cycle, and up to 60% of the mass flux during times of maximum activity (*23*). Because plumes fill only 10% of a coronal hole's volume (*25*), the bulk of the wind might originate in inter-plume regions (*26–28*). However, the spatial structure and timescales of inter-plume outflows are not known.

The solar wind emerging from coronal holes connects the Sun to the heliosphere. Spacecraft measurements in the inner heliosphere have been interpreted as showing that reversals of the radial magnetic field (termed switchbacks) in the solar wind are modulated on spatial scales as small as 1 Mm (typical photospheric convective granular scale), at the base of the Sun's corona (*29*). However, EUV imaging has not been available with sub-granular resolution (< 1 Mm), which is necessary to investigate this proposal.

**EUV observations of a coronal hole**

We observed a coronal hole, located near the Sun's south pole, using the EUV High Resolution Imager (HRI$_{EUV}$), which is part of the Extreme Ultraviolet Imager (EUI) (*30*) instrument on the Solar Orbiter spacecraft (*31*). The data were obtained during a perihelion observing campaign of Solar Orbiter on 2022 March 30 between 04:30 and 05:00 Universal Time (UT). The images have a spatial resolution of about 237 km on the Sun and a cadence of 3 s (*32*). The passband of HRI$_{EUV}$ is centered at 17.4 nm, with its thermal response function peaking around temperatures of about 1 MK due to emission lines of Fe ɪx (at 17.11 nm) and Fe x (at 17.45 nm and 17.72 nm).

We split the coronal hole into three regions (labeled R1–R3 in Fig. 1A) to investigate the small-scale coronal dynamics. Regions R1 and R3 are darker portions of the coronal hole with a few embedded brighter emission structures known as coronal bright points, while R2 is an isolated plume within the coronal hole.

**Y-shaped transient jets**

We identify a Y-shaped jet with a base extent of about 5 Mm between boxes R1-2 and R1-4 in Fig. 1B. Jets with such Y-shaped morphology are common in coronal holes (*33*). We focus on smaller jets, on spatial scales<1 Mm, emerging from the ambient coronal hole. We visually identified 10 such small-scale jets within R1 (Fig. 1C–L). Some of these jets show a full or partial Y-shaped morphology (jets 1, 2, 3, 6, 7, 8, and 10), while some show a much simpler linear morphology (jets 4, 5, and 9). It is possible that all are intrinsically Y-shaped, but this is not visible in some cases because of the viewing angle, which could lead to both legs of the jet being aligned with the line-of-sight. All these jets have fine-scale structure down to the spatial resolution of our images.

These small jets exhibit intensity enhancements over the background emission from the coronal hole. Fig. 2 shows light curves of three example jets in region R1. For each of these, we selected two spatial points along the path of the jet to study the intensity variations (Fig. S1). The intensity enhancements (defined as the amplitude of intensity fluctuations compared to the local background) are in the range of 100 to 300 digital numbers (DN) pixel$^{-1}$ (equivalent to counts of about 15 to 44 photons pixel$^{-1}$; for comparison, coronal bright points in our observations are ~500 photons pixel$^{-1}$). Depending on the number of pixels used for spatial averaging of the signal, peak intensities of all these jets are at least 3 times the local background photon noise (*32*) above the background average intensity.

The observed intensity fluctuations vary on timescales of 20 to 60 s (Fig. 2). Features associated with an intense jet-base (Fig.1E–H), at least partially persist for longer timescales of 300 to 600 s, but also exhibit intensity fluctuations on shorter timescales of 100 s. We estimate that the jet speeds are ~ 100 km s$^{-1}$, within a factor of two (Fig. 2D-F) (*32*). These differences in measured jet speeds could be due to projection effects. We observed repeated jet activity in these regions and visually counted a total of 38 small jets from the 10 locations labelled in Fig. 1 over the course of 30 minutes of observations. We also found 29 additional small-scale jets in the wider coronal hole region (Movie S1).

**Plume-related jets**

The emission from region R2 (Fig. 1A) is enhanced over the immediate surroundings, which suggests that it is a coronal hole plume. We observed clusters of small jets emerging throughout the associated plume region (Fig. 3). These plume jets do not have the Y-shaped morphology of the jets discussed above. Previous lower resolution EUV observations have shown that outflows from plumes are structured in the form of jets (*21*) and that the outflows are quasi-periodic in nature, with speeds of ~100 km s$^{-1}$ (*19, 23*). Our higher resolution HRI$_{EUV}$ observations show even smaller jets, with cross-sectional widths close to the spatial resolution limit of ~200 km (a very thin jet is shown in Fig. 3H). We observe at least 30 such jets from the plume (Movie S2).

**Faint inter-plume emission and jets**

While both the inter-plume Y-shaped jets and the plume jets are spatially discrete, we found that there is a more widespread emission that is faint, yet above the background level. This fainter emission, with veil-like morphology, exhibits apparent outward propagation throughout the coronal hole (Fig. S3; Movies S3 and S4). The fainter emission also originates from the inter-plume regions; we designate four examples as R1-a, R1-b, R1-c and R3-a (Fig. 1A). These regions of fainter emission are partially obscured by the foreground emission in R1, but the source region in R3 is unobscured.

This fainter veil-like emission is also spatially structured. There are outflows apparently emerging from the identified source regions, in the form of small-scale jets. Fig. 4 shows three example jets from three of the identified inter-plume sources. Compared to the Y-shaped jets in Fig. 1, these inter-plume jets are even fainter; though at least 3 times the local background photon noise above the background mean intensities. These fainter jets exhibit intensity enhancements of about 50 DN pixel$^{-1}$ over the local background (Figs. 4 and S2). These inter-plume jets have similar morphology to the plume counterparts and their cross-sectional widths are close to the resolution limit of the images. Inter-plume jets also appear in clusters on larger-scales, > 10 Mm, similar to plume jets. They are associated with intensity propagations; although we cannot determine these jet speeds due to noise, we constrain them to be supersonic outflows > 100 km s$^{-1}$ (*32*).

**Magnetic origins of the jets**

The high speeds, short timescales and narrow widths of the observed jets indicate that they are driven by a variable energy source, located at the coronal base, on spatial scales of ≲ 100 km. Y-shaped jets (Fig. 1) are already known to be driven by reconnection of open and closed magnetic field lines (*34–36*). The base-width of ≲ 1 Mm indicates that the observed small Y-shaped jets are associated with emergence or cancellation of granular-scale magnetic features (*37*). Although the plume and fainter inter-plume jets we observe on the solar disc do not have Y-shaped morphology, similar features observed on the Sun's limb do have morphologies consistent with magnetic topology of interchange reconnection between open and closed magnetic domains (Figs. S4–S6; Movies S5–S9). We therefore suggest that all the jets we observed are transiently driven by reconnection.

Previous observations have shown that the Ne VIII emission line, which samples transition region and coronal plasma around and above 0.63 MK, has systematic blueshifts indicating upwards flows of about 10 km s$^{-1}$ in coronal holes (*13, 38*). The outflow speeds in inter-plume regions can be up to 70 km s$^{-1}$ (*27, 28*). The small-scale jets we observed could be the ~1 MK counterparts of the coronal hole outflows previously inferred through spectroscopy at lower spatial resolution. These high-speed jets driven by reconnection in our observations propagate outward and could channel some, if not all, of the material contained within the jets, along the open magnetic field lines of the coronal hole, powering the solar wind.

We do not observe any signatures of downwards flows in the jets. If they exist, they could be too weak to be detectable in our observations, or might contain plasma at different temperatures that the HRI$_{EUV}$ instrument is not sensitive to.

**Jet energetics**

Assuming that the apparent outflows we detected are mass motions, we estimate a lower limit on the kinetic energy flux of ≳ 10$^5$ erg cm$^{-2}$ s$^{-1}$ (*32*). This is probably an under-estimate due to line-of-sight effects. Increasing the jet speed by a factor of 2 to 4 would increase the kinetic energy flux of a given jet by nearly one to two orders of magnitude.

Our estimate of the kinetic energy flux in these jets is similar to the minimum non-radiative energy input required to power magnetically open coronal regions (*39*). The narrowest jets we observe have cross-sectional widths of only a few pixels, corresponding to diameters of < 237 km, assuming a cylindrical shape. The observed intensity fluctuations are on timescales of 20 to 100 s. If we take that to be the lifetime of the jet, and combine it with the derived kinetic energy flux, we conclude that the lower limit for the kinetic energy content of a single narrow, faint jet is ≳10$^{21}$ erg. The typical energy content of a nanoflare is 10$^{24}$ erg (*40*), so the kinetic energy content of fainter inter-plume jets is in the picoflare range, so we refer to them as picoflare jets. The Y-shaped jets (Fig. 1) and the plume jets (Fig. 3), which are somewhat brighter, presumably contain more kinetic energy.

Plasma outflows from picoflare jets, channeled along the open magnetic field lines of coronal holes, could contribute to the mass flux of the solar wind. We estimate the total mass-loss rate, $\dot{m}$, of the observed jets as $\dot{m} = 4\pi R_\odot^2 \rho v f_{\mathrm{CH}} f_{\mathrm{j}}$, where, $R_\odot = 695$ Mm is the radius of the Sun, $\rho = 3.34 \times 10^{-16}$ g cm$^{-3}$ is the jet mass density, $v = 100$ km s$^{-1}$ is the jet speed (*32*), $f_{\mathrm{CH}} \approx 0.1$ is the fractional area coverage of coronal holes on the Sun (*41*) and $f_{\mathrm{j}}$ is the filling factor of picoflare jets. During periods of high solar activity, upwards flows from the edges of active regions could also contribute to the solar wind mass flux (*42–45*). If picoflare jets account for the wind mass flux of ~$10^{12}$ g s$^{-1}$ throughout the solar cycle (*46*), their filling factor $f_{\mathrm{j}}$ would need to be ~0.05.

Our observations do not spatially resolve the smallest jets (*32*), especially those in the inter-plume regions which have weaker intensity enhancements and lower contrast with respect to the local background. Our results are consistent with previous EUV imaging at lower spatial resolution (*24*). The picoflare jets are common in our images (Figs. S7–S8), particularly in the inter-plume regions that cover a substantial portion of the coronal hole (Movies S10–S12). Above the limb, we observe persistent jet activity that fills the plane-of-sky (Fig. S9). Any fainter jets would not be distinguishable from the background intensity fluctuations in our data.

**Implications for the solar wind**

Given their frequency in our observations, we suggest these picoflare jets might be ubiquitous. If so, those emerging from the inter-plume regions, and to a lesser extent those from plume regions, could provide substantial plasma and energy flux to the solar wind, throughout the solar cycle. This is consistent with previous suggestions that small jets make a substantial contribution to the solar wind (*24*).

The intermittent variability and short lifetimes (20 to 100 s) of the observed jets are consistent with the inherently ephemeral nature of magnetic reconnection. Our observations are not sensitive to any steady component of the solar wind. Nevertheless, we suggest that a dynamic component of the solar wind emerges from coronal holes as an intermittent outflow at the coronal base, which is observed as picoflare jets.

Neighboring small-scale high-speed outflows with different speeds would produce a velocity shear, leading to instabilities (*47, 48*). We suggest this could occur in both plume and inter-plume regions (*32*). If picoflare jets are ubiquitous, these instabilities could play a role in the formation of structures in the solar wind, such as magnetic switchbacks (*7*). In situ spacecraft measurements of the solar wind in the inner heliosphere have been interpreted as showing that the sources responsible for the large- and mid-scale modulation of switchbacks operate on spatial scales, comparable to typical convective scales of supergranules to granules on the Sun (*29*). The narrow widths of the picoflare jets we observe is compatible with them having underlying source regions <1 Mm, i.e., smaller than the typical granular scale (*49*). The larger clusters of picoflare jets we observe have structure on scales > 10 Mm, which is close to the supergranular scales of 10 to 60 Mm (*50*). Our interpretation that the picoflare jets are driven by granular-scale reconnection, could therefore be related to the mid- to large-scale modulations of switchbacks in the solar wind.

The lifetimes and speeds of the picoflare jets are similar to those of network jets, which are common in the transition region (*16*), so the phenomena might be related. Our data cannot determine whether a hot picoflare jet (~1 MK) hosts plasma that has been heated from a cooler network jet phase, or if there are two populations of jets present simultaneously (colder network jets and hotter coronal picoflare jets). We suggest that picoflare jets could supply heated material to the solar wind, as could the transition region jets if their energy is dissipated (*16*).

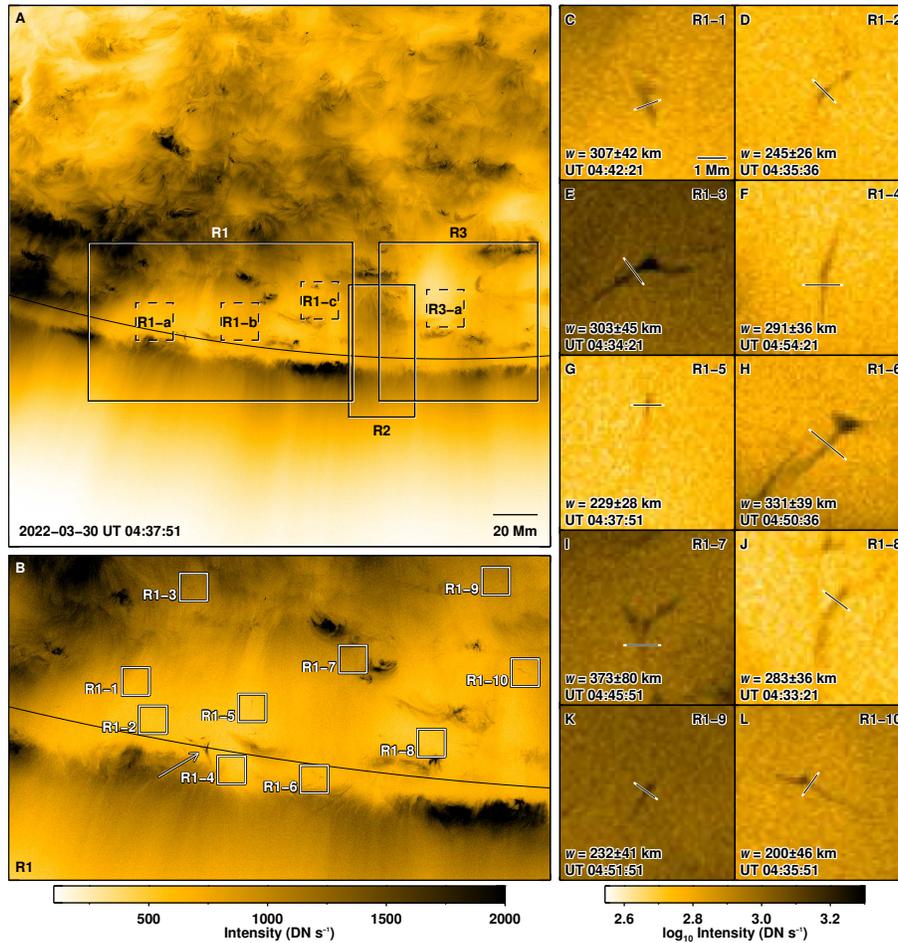

**Fig. 1. EUV images of jets from a coronal hole**. All images are displayed on a negative colour scale (i.e., black is bright). Each panel shows data with a 15 s effective cadence (*32*) and south is roughly downwards. (**A**) An overview HRI$_{EUV}$ image of a south polar coronal hole recorded on 2022 March 30 around 04:38 UT. The field of view is about 243 Mm × 243 Mm. Solid boxes labeled R1–R3 outline regions of the coronal hole discussed in the text. Dashed boxes labeled R1-a, R1-b, R1-c and R3-a are inter-plume regions. The solid curve indicates the position of the Sun's limb. (**B**) Zoomed-in view of R1 from panel **A**, on the same brightness scale. Boxes labeled R1-1 to R1-10, each about 6 Mm × 6 Mm, indicate locations within the coronal hole where we identified small-scale jet activity. The arrow points to a ~ 5 Mm scale larger Y-shaped jet. An animated version of this panel is shown in Movie S1. (**C**–**L**) Closer views of the boxes in panel B, on a logarithmic brightness scale. White lines indicate locations where the width of each jet, *w*, was measured; its value is shown at the bottom of each panel (also listed in Table S1).

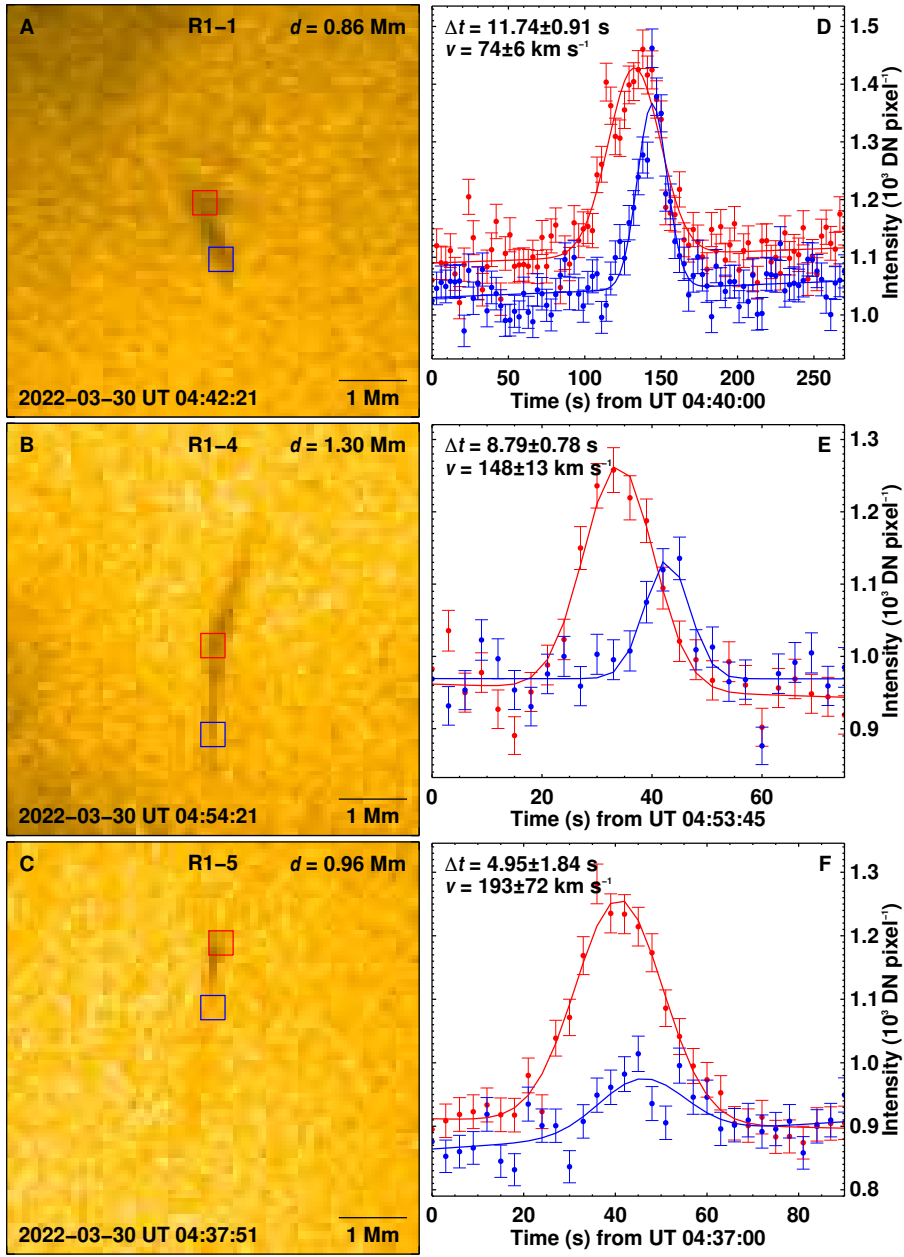

**Fig. 2. Temporal variability in three example small-scale jets. (A-C)** Same EUV images as Fig. 1C, F and G respectively. The red and blue squares mark locations along each jet, separated by a plane-of-sky distance, $d$, listed in the top right of each panel. **(D-F)** Average intensities (data points) within the corresponding red and blue boxes in panels A-C as a function of time, with a cadence of 3 s. Error bars indicate $1\sigma$ uncertainties because of photon Poisson (shot) noise (*32*). The red and blue solid curves are Gaussian models fitted to the observed intensities. The separation between the centroids of the two Gaussian peaks, $\Delta t$, and the speed, $v = d/\Delta t$ with their standard errors are listed in each panel (also listed in Table S2).

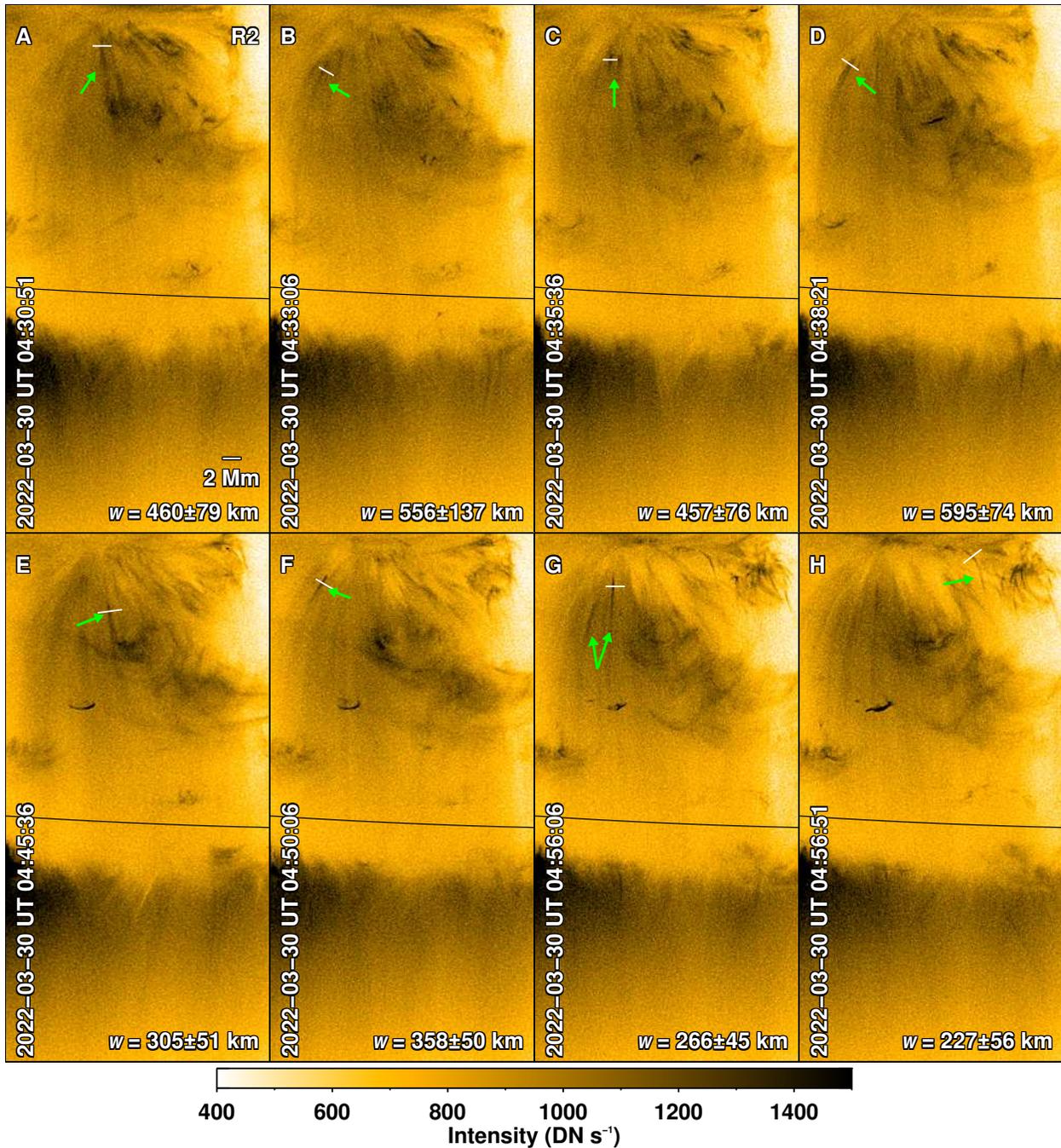

**Fig. 3. Jets from plumes.** The panels all show region R2 from Fig. 1A, but for different times (labelled on each image) with an effective cadence of 15 s (*32*). Small-scale jet activity is indicated by green arrows. White lines indicate locations where the cross-sectional width, $w$, of each jet was measured, with the results listed in each panel (also listed in Table S3). The solid curve marks the solar limb. An animated version of this Figure is shown in Movie S2.

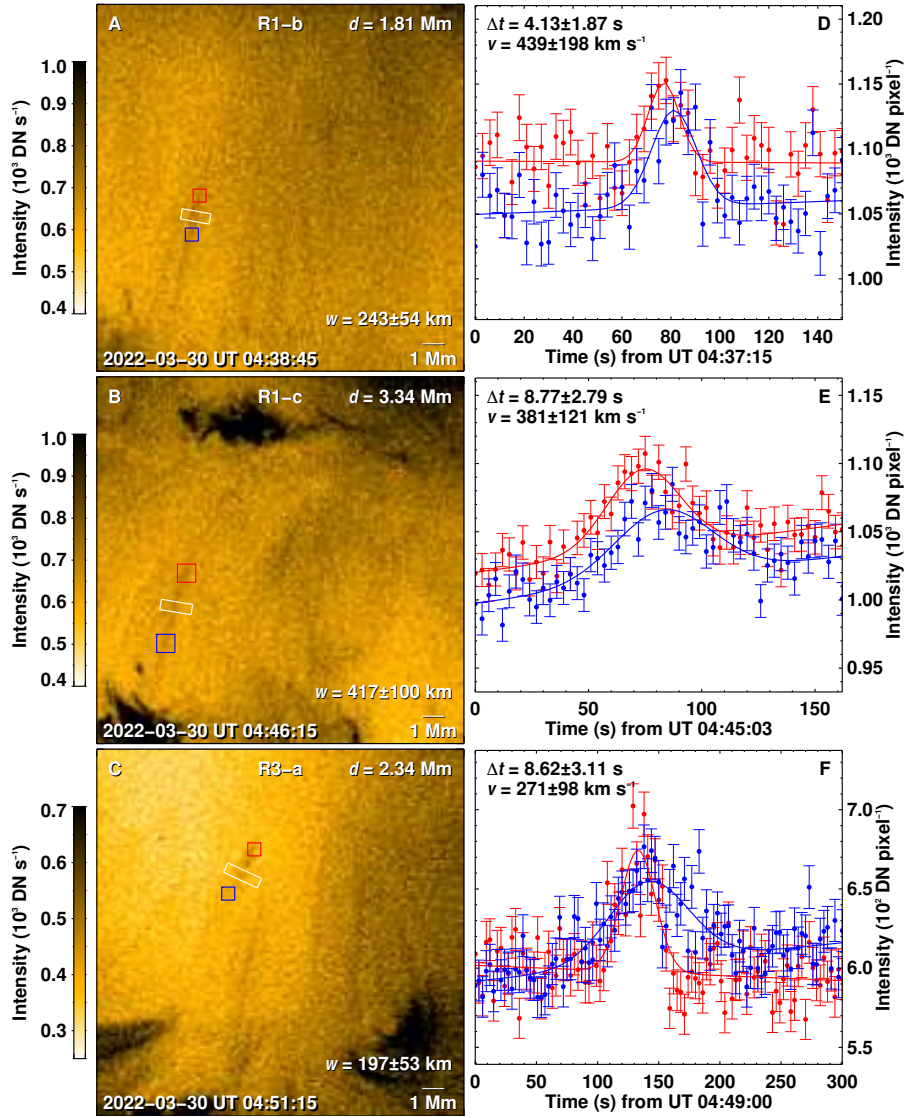

**Fig. 4. High-speed faint jets from inter-plume regions.** Same as Fig. 2, but for the boxes labelled R1-b, R1-c, and R3-a in Fig. 1A. Cross-sectional width, *w*, of each jet was measured by averaging the intensity across the white strip, with the results listed in each panel. The images in panels A-C have an effective cadence of 30 s. The average intensities in panels D-F have a cadence of 3 s (*32*). Jet parameters also listed in Table S4.

*Science* **381**, 867–872 (2023). DOI: [10.1126/science.ade5801](10.1126/science.ade5801)

**Acknowledgments:** Solar Orbiter is a space mission with international collaboration between ESA and NASA, operated by ESA. The EUI instrument was built by CSL, IAS, MPS, MSSL/UCL, PMOD/WRC, ROB, LCF/IO with funding from the Belgian Federal Science Policy Office (BELSPO/PRODEX PEA 4000134088); the Centre National d'Etudes Spatiales (CNES); the UK Space Agency (UKSA); the Bundesministerium für Wirtschaft und Energie (BMWi) through the Deutsches Zentrum für Luft- und Raumfahrt (DLR); and the Swiss Space Office (SSO). This research has made use of NASA's Astrophysics Data System. **Funding:** LPC acknowledges funding by the European Union (ERC, ORIGIN, 101039844). SP acknowledges funding by CNES through the MEDOC data and operations center. DML thanks the Science and Technology Facilities Council for the award of an Ernest Rutherford Fellowship (ST/R003246/1). ANZ, DB, EK, LR, and CV thank the Belgian Federal Science Policy Office (BELSPO) for the provision of financial support in the framework of the PRODEX Programme of the European Space Agency (ESA) under contract numbers 4000134474 and 4000136424. **Author contributions:** LPC led the study, data analysis and wrote the manuscript with inputs from ANZ, DB, HP, SP, SM, RAC, US, LT, ÉB, DML, and DBS. FA, EK, CV contributed to data reduction. ANZ led the Solar Orbiter observing campaign. DB is the Principal Investigator of EUI. All the authors discussed and interpreted the results. **Competing interests:** There are no competing interests to declare. **Data and materials availability:** The $HRI_{EUV}$ level-2 data are archived by the Royal Observatory of Belgium (*51*). We used observations in the time range 2022 March 30 04:30–05:00 UT (files from solo_L2_eui-hrieuvnon-image_20220330T043000227_V01.fits to solo_L2_eui-hrieuvnon-image_20220330T045957224_V01.fits). These data were recorded as a part of the Solar Orbiter Observing Plan (SOOP), R_S<small>MALL</small>_H<small>RES</small>_M<small>CAD</small>_P<small>OLAR</small>-O<small>BSERVATIONS</small> (*52*). Using the SOOP name and the time range mentioned above, these data can be retrieved at ESA's Solar Orbiter Archive (https://soar.esac.esa.int/soar/).

**References**

1. E. N. Parker, Dynamics of the Interplanetary Gas and Magnetic Fields. *Astrophys. J.* **128**, 664 (1958).
2. M. Neugebauer, C. W. Snyder, Solar Plasma Experiment. *Science*. **138**, 1095–1097 (1962).
3. J. Woch, W. I. Axford, U. Mall, B. Wilken, S. Livi, J. Geiss, *et al.*, SWICS/Ulysses observations: The three-dimensional structure of the heliosphere in the declining/minimum phase of the solar cycle. *Geophys. Res. Lett*. **24**, 2885–2888 (1997).
4. D. J. McComas, B. L. Barraclough, H. O. Funsten, J. T. Gosling, E. Santiago-Muñoz, R. M. Skoug, *et al.*, Solar wind observations over Ulysses' first full polar orbit. *J. Geophys. Res*. **105**, 10419–10434 (2000).
5. S. R. Cranmer, Coronal Holes. *Living Rev. Sol. Phys*. **6**, 3 (2009).
6. Y.-M. Wang, Y.-K. Ko, Observations of Slow Solar Wind from Equatorial Coronal Holes. *Astrophys. J.* **880**, 146 (2019).
7. S. D. Bale, S. T. Badman, J. W. Bonnell, T. A. Bowen, D. Burgess, A. W. Case, *et al.*, Highly structured slow solar wind emerging from an equatorial coronal hole. *Nature*. **576**, 237–242 (2019).
8. G. Poletto, Solar Coronal Plumes. *Living Rev. Sol. Phys*. **12**, 7 (2015).
9. N. M. Viall, J. E. Borovsky, Nine Outstanding Questions of Solar Wind Physics. *J. Geophys. Res. Space Phys*. **125**, e26005 (2020).

10. S. R. Cranmer, A. A. van Ballegooijen, R. J. Edgar, Self-consistent Coronal Heating and Solar Wind Acceleration from Anisotropic Magnetohydrodynamic Turbulence. *Astrophys. J. Suppl. Ser.* **171**, 520–551 (2007).
11. T. Matsumoto, T. K. Suzuki, Connecting the Sun and the Solar Wind: The First 2.5-dimensional Self-consistent MHD Simulation under the Alfvén Wave Scenario. *Astrophys. J.* **749**, 8 (2012).
12. M. Shoda, B. D. G. Chandran, S. R. Cranmer, Turbulent Generation of Magnetic Switchbacks in the Alfvénic Solar Wind. *Astrophys. J.* **915**, 52 (2021).
13. C.-Y. Tu, C. Zhou, E. Marsch, L.-D. Xia, L. Zhao, J.-X. Wang, *et al.*, Solar Wind Origin in Coronal Funnels. *Science*. **308**, 519–523 (2005).
14. L. Yang, J. He, H. Peter, C. Tu, W. Chen, L. Zhang, *et al.*, Injection of Plasma into the Nascent Solar Wind via Reconnection Driven by Supergranular Advection. *Astrophys. J.* **770**, 6 (2013).
15. Y.-M. Wang, Small-scale Flux Emergence, Coronal Hole Heating, and Flux-tube Expansion: A Hybrid Solar Wind Model. *Astrophys. J.* **904**, 199 (2020).
16. H. Tian, E. E. DeLuca, S. R. Cranmer, B. De Pontieu, H. Peter, J. Martínez-Sykora, *et al.*, Prevalence of small-scale jets from the networks of the solar transition region and chromosphere. *Science*. **346**, 1255711 (2014).
17. P. Kayshap, K. Murawski, A. K. Srivastava, B. N. Dwivedi, Rotating network jets in the quiet Sun as observed by IRIS. *Astron. Astrophys.* **616**, A99 (2018).
18. J. Gorman, L. P. Chitta, H. Peter, Spectroscopic observation of a transition region network jet. *Astron. Astrophys.* **660**, A116 (2022).
19. H. Tian, S. W. McIntosh, S. R. Habbal, J. He, Observation of High-speed Outflow on Plume-like Structures of the Quiet Sun and Coronal Holes with Solar Dynamics Observatory/Atmospheric Imaging Assembly. *Astrophys. J.* **736**, 130 (2011).
20. S. Pucci, G. Poletto, A. C. Sterling, M. Romoli, Birth, Life, and Death of a Solar Coronal Plume. *Astrophys. J.* **793**, 86 (2014).
21. N.-E. Raouafi, G. Stenborg, Role of Transients in the Sustainability of Solar Coronal Plumes. *Astrophys. J.* **787**, 118 (2014).
22. V. M. Uritsky, C. E. DeForest, J. T. Karpen, C. R. DeVore, P. Kumar, N. E. Raouafi, *et al.*, Plumelets: Dynamic Filamentary Structures in Solar Coronal Plumes. *Astrophys. J.* **907**, 1 (2021).
23. P. Kumar, J. T. Karpen, V. M. Uritsky, C. E. Deforest, N. E. Raouafi, C. Richard DeVore, Quasi-periodic Energy Release and Jets at the Base of Solar Coronal Plumes. *Astrophys. J.* **933**, 21 (2022).
24. N. E. Raouafi, G. Stenborg, D. B. Seaton, H. Wang, J. Wang, C. E. DeForest, *et al.*, Magnetic Reconnection as the Driver of the Solar Wind. *Astrophys. J.* **945**, 28 (2023).
25. I. A. Ahmad, G. L. Withbroe, EUV analysis of polar plumes. *Sol. Phys.* **53**, 397–408 (1977).
26. Y.-M. Wang, Polar Plumes and the Solar Wind. *Astrophys. J. Lett.* **435**, L153 (1994).
27. S. Patsourakos, J.-C. Vial, Outflow velocity of interplume regions at the base of Polar Coronal Holes. *Astron. Astrophys.* **359**, L1–L4 (2000).
28. L. Teriaca, G. Poletto, M. Romoli, D. A. Biesecker, The Nascent Solar Wind: Origin and Acceleration. *Astrophys. J.* **588**, 566–577 (2003).

29. N. Fargette, B. Lavraud, A. P. Rouillard, V. Réville, T. Dudok De Wit, C. Froment, *et al.*, Characteristic Scales of Magnetic Switchback Patches Near the Sun and Their Possible Association With Solar Supergranulation and Granulation. *Astrophys. J.* **919**, 96 (2021).
30. P. Rochus, F. Auchère, D. Berghmans, L. Harra, W. Schmutz, U. Schühle, *et al.*, The Solar Orbiter EUI instrument: The Extreme Ultraviolet Imager. *Astron. Astrophys.* **642**, A8 (2020).
31. D. Müller, O. C. St. Cyr, I. Zouganelis, H. R. Gilbert, R. Marsden, T. Nieves-Chinchilla, *et al.*, The Solar Orbiter mission. Science overview. *Astron. Astrophys.* **642**, A1 (2020).
32. Materials and methods are available as supplementary materials.
33. F. Moreno-Insertis, K. Galsgaard, I. Ugarte-Urra, Jets in Coronal Holes: Hinode Observations and Three-dimensional Computer Modeling. *Astrophys. J. Lett.* **673**, L211 (2008).
34. K. Shibata, T. Nakamura, T. Matsumoto, K. Otsuji, T. J. Okamoto, N. Nishizuka, *et al.*, Chromospheric Anemone Jets as Evidence of Ubiquitous Reconnection. *Science*. **318**, 1591 (2007).
35. A. C. Sterling, R. L. Moore, D. A. Falconer, M. Adams, Small-scale filament eruptions as the driver of X-ray jets in solar coronal holes. *Nature*. **523**, 437–440 (2015).
36. S. Mandal, L. P. Chitta, H. Peter, S. K. Solanki, R. A. Cuadrado, L. Teriaca, *et al.*, A highly dynamic small-scale jet in a polar coronal hole. *Astron. Astrophys.* **664**, A28 (2022).
37. L. P. Chitta, A. R. C. Sukarmadji, L. Rouppe van der Voort, H. Peter, Energetics of magnetic transients in a solar active region plage. *Astron. Astrophys.* **623**, A176 (2019).
38. L. D. Xia, E. Marsch, W. Curdt, On the outflow in an equatorial coronal hole. *Astron. Astrophys.* **399**, L5–L9 (2003).
39. G. L. Withbroe, The Temperature Structure, Mass, and Energy Flow in the Corona and Inner Solar Wind. *Astrophys. J.* **325**, 442 (1988).
40. E. N. Parker, Nanoflares and the Solar X-Ray Corona. *Astrophys. J.* **330**, 474 (1988).
41. K. L. Harvey, F. Recely, Polar Coronal Holes During Cycles 22 and 23. *Sol. Phys.* **211**, 31–52 (2002).
42. T. Sakao, R. Kano, N. Narukage, J. Kotoku, T. Bando, E. E. DeLuca, *et al.*, Continuous Plasma Outflows from the Edge of a Solar Active Region as a Possible Source of Solar Wind. *Science*. **318**, 1585 (2007).
43. G. A. Doschek, J. T. Mariska, H. P. Warren, C. M. Brown, J. L. Culhane, H. Hara, *et al.*, Nonthermal Velocities in Solar Active Regions Observed with the Extreme-Ultraviolet Imaging Spectrometer on Hinode. *Astrophys. J. Lett.* **667**, L109–L112 (2007).
44. L. K. Harra, T. Sakao, C. H. Mandrini, H. Hara, S. Imada, P. R. Young, *et al.*, Outflows at the Edges of Active Regions: Contribution to Solar Wind Formation? *Astrophys. J. Lett.* **676**, L147 (2008).
45. D. H. Brooks, L. Harra, S. D. Bale, K. Barczynski, C. Mandrini, V. Polito, *et al.*, The Formation and Lifetime of Outflows in a Solar Active Region. *Astrophys. J.* **917**, 25 (2021).
46. Y.-M. Wang, in *Cool Stars, Stellar Systems, and the Sun*, R. A. Donahue, J. A. Bookbinder, Eds. (1998), vol. **154** of *Astronomical Society of the Pacific Conference Series*, p. 131.
47. S. Parhi, S. T. Suess, M. Sulkanen, Can Kelvin-Helmholtz instabilities of jet-like structures and plumes cause solar wind fluctuations at 1 AU? *J. Geophys. Res*. **104**, 14781–14788 (1999).

48. J. Andries, M. Goossens, Kelvin-Helmholtz instabilities and resonant flow instabilities for a coronal plume model with plasma pressure. *Astron. Astrophys.* **368**, 1083–1094 (2001).
49. Th. Roudier, R. Muller, Structure of the solar granulation. *Sol. Phys.* **107**, 11–26 (1986).
50. G. W. Simon, N. O. Weiss, Supergranules and the Hydrogen Convection Zone. *Zeitschrift für Astrophysik*. **69**, 435 (1968).
51. B. Mampaey, F. Verbeeck, K. Stegen, E. Kraaikamp, S. Gissot, F. Auchère, *et al.*, SolO/EUI Data Release 5.0 2022-04, Royal Observatory of Belgium (2022) DOI:10.24414/2qfw-tr95.
52. I. Zouganelis, A. De Groof, A. P. Walsh, D. R. Williams, D. Müller, O. C. St Cyr, *et al.*, The Solar Orbiter Science Activity Plan. Translating solar and heliospheric physics questions into action. *Astron. Astrophys.* **642**, A3 (2020).
53. L. P. Chitta, H. Peter, S. Parenti, D. Berghmans, F. Auchère, S. K. Solanki, *et al.*, Solar coronal heating from small-scale magnetic braids. *Astron. Astrophys.* **667**, A166 (2022).
54. C. B. Markwardt, in *Astronomical Data Analysis Software and Systems XVIII.* D. A. Bohlender, D. Durand, P. Dowler, Eds. (2009), vol. **411** of *Astronomical Society of the Pacific Conference Series*, p. 251.
55. S. Gissot, F. Auchère, D. Berghmans, B. Giordanengo, A. BenMoussa, J. Rebellato, *et al.*, Initial radiometric calibration of the High-Resolution EUV Imager (HRI$_{EUV}$) of the Extreme Ultraviolet Imager (EUI) instrument onboard Solar Orbiter. *arXiv e-print 2307.14182* (2023) DOI:10.48550/arXiv.2307.14182.
56. J.-L. Starck, F. Murtagh, Image restoration with noise suppression using the wavelet transform. *Astron. Astrophys.* **288**, 342–348 (1994).
57. G. A. Doschek, H. P. Warren, J. M. Laming, J. T. Mariska, K. Wilhelm, P. Lemaire, *et al.*, Electron Densities in the Solar Polar Coronal Holes from Density-Sensitive Line Ratios of Si VIII and S X. *Astrophys. J. Lett.* **482**, L109–L112 (1997).
58. K. Wilhelm, Solar coronal-hole plasma densities and temperatures. *Astron. Astrophys.* **455**, 697–708 (2006).
59. H. N. Smitha, L. S. Anusha, S. K. Solanki, T. L. Riethmüller, Estimation of the Magnetic Flux Emergence Rate in the Quiet Sun from Sunrise Data. *Astrophys. J. Suppl. Ser.* **229**, 17 (2017).
60. L. P. Chitta, H. Peter, S. K. Solanki, Nature of the energy source powering solar coronal loops driven by nanoflares. *Astron. Astrophys.* **615**, L9 (2018).
61. E. R. Priest, L. P. Chitta, P. Syntelis, A Cancellation Nanoflare Model for Solar Chromospheric and Coronal Heating. *Astrophys. J. Lett.* **862**, L24 (2018).
62. V. Upendran, D. Tripathi, On the Formation of Solar Wind and Switchbacks, and Quiet Sun Heating. *Astrophys. J.* **926**, 138 (2022).
63. S. R. Cranmer, Low-frequency Alfvén Waves Produced by Magnetic Reconnection in the Sun's Magnetic Carpet. *Astrophys. J.* **862**, 6 (2018).
64. E. Marsch, Kinetic Physics of the Solar Corona and Solar Wind. *Living Rev. Sol. Phys*. **3**, 1 (2006).
65. A. R. Paraschiv, A. Bemporad, A. C. Sterling, Physical properties of solar polar jets. A statistical study with Hinode XRT data. *Astron. Astrophys.* **579**, A96 (2015).
66. L. P. Chitta, D. B. Seaton, C. Downs, C. E. DeForest, A. K. Higginson, Direct observations of a complex coronal web driving highly structured slow solar wind. *Nat Astron*. **7**, 133–141 (2023).

67. G. R. Gupta, L. Teriaca, E. Marsch, S. K. Solanki, D. Banerjee, Spectroscopic observations of propagating disturbances in a polar coronal hole: evidence of slow magneto-acoustic waves. *Astron. Astrophys.* **546**, A93 (2012).
68. S. D. Bale, J. F. Drake, M. D. McManus, M. I. Desai, S. T. Badman, D. E. Larson, *et al.*, Interchange reconnection as the source of the fast solar wind within coronal holes. *Nature*. **618**, 252–256 (2023).
69. T. Van Doorsselaere, N. Wardle, G. Del Zanna, K. Jansari, E. Verwichte, V. M. Nakariakov, The First Measurement of the Adiabatic Index in the Solar Corona Using Time-dependent Spectroscopy of Hinode/EIS Observations. *Astrophys. J. Lett.* **727**, L32 (2011).
70. J. W. Cirtain, L. Golub, L. Lundquist, A. van Ballegooijen, A. Savcheva, M. Shimojo, *et al.*, Evidence for Alfvén Waves in Solar X-ray Jets. *Science*. **318**, 1580 (2007).
71. U. V. Möstl, M. Temmer, A. M. Veronig, The Kelvin-Helmholtz Instability at Coronal Mass Ejection Boundaries in the Solar Corona: Observations and 2.5D MHD Simulations. *Astrophys. J. Lett.* **766**, L12 (2013).
72. L. P. Chitta, E. R. Priest, X. Cheng, From Formation to Disruption: Observing the Multiphase Evolution of a Solar Flare Current Sheet. *Astrophys. J.* **911**, 133 (2021).
73. X. Li, J. Zhang, S. Yang, Y. Hou, R. Erdélyi, Observing Kelvin-Helmholtz instability in solar blowout jet. *Sci Rep*. **8**, 8136 (2018).
74. P. Antolin, T. J. Okamoto, B. De Pontieu, H. Uitenbroek, T. Van Doorsselaere, T. Yokoyama, Resonant Absorption of Transverse Oscillations and Associated Heating in a Solar Prominence. II. Numerical Aspects. *Astrophys. J.* **809**, 72 (2015).
75. T. A. Howson, I. De Moortel, P. Antolin, Energetics of the Kelvin-Helmholtz instability induced by transverse waves in twisted coronal loops. *Astron. Astrophys.* **607**, A77 (2017).

**Supplementary Materials:**

**Materials and Methods**
**Supplementary Text**
**Figs. S1 to S9**
**Tables S1 to S4**
**Captions for Movies S1 to S12**
**References (53–75)**

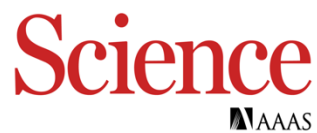

# Supplementary Materials for

## Picoflare jets power the solar wind emerging from a coronal hole on the Sun

L. P. Chitta[*], A. N. Zhukov, D. Berghmans, H. Peter, S. Parenti, S. Mandal,
R. Aznar Cuadrado, U. Schühle, L. Teriaca, F. Auchère, K. Barczynski, É. Buchlin,
L. Harra, E. Kraaikamp, D. M. Long, L. Rodriguez, C. Schwanitz, P. J. Smith, C. Verbeeck,
D. B. Seaton

*Corresponding author. Email: chitta@mps.mpg.de

**This PDF file includes:**

- Materials and Methods
- Supplementary Text
- Figures S1 to S9
- Tables S1 to S4
- Captions for Movies S1 to S12

**Other supplementary materials:**

- Movies S1 to S12

## Materials and Methods

We used EUV images of the south polar coronal hole recorded by the HRI$_{EUV}$ instrument. During the observing campaign of Solar Orbiter on 2022 March 30 between 04:30 and 05:00 Universal Time (UT), the spacecraft was at a distance of 0.332 astronomical units (au) from the Sun. The data were recorded at a cadence of 3 s, with an exposure time of 1.65 s, and the images have a plate scale of 0.492 arcseconds (″) pixel$^{-1}$. The spatial resolution of HRI$_{EUV}$ at the time of the observations is about 237 km on the Sun (the full width at half maximum of the point spread function of HRI$_{EUV}$ is about two pixels). We used level-2 data from this observing campaign (*51*).

The jitter in the level-2 HRI$_{EUV}$ data was removed as follows. We divided the whole observational sequence into shorter overlapping sequences, each containing 30 images. The overlap is such that the last image from a given shorter sequence is the same as the first image from the following shorter sequence. Then through cross-correlation, we aligned each image to the first image in a given shorter sequence. Because the shorter sequences overlap, this resulted in a dataset in which all images from the whole observational sequence were aligned to the first image of the time series (*53*).

We focus on the EUV jet features with intensity enhancements in the range of 0.01 to 0.1 times the intensity of quiet-Sun features, such as coronal bright points observed in the same data. To visualize the fainter features (only for display purposes) we further processed the level-2 data as follows. First, to improve the signal from the fainter features, we temporally binned the data collected over 15 s non-overlapping intervals. This amounts to averaging every five images (e.g., images 1–5, 6–10, 11–15, and so on). This results in a dataset with an effective cadence of 15 s. In this sequence, the time stamps are obtained from every third image from the original 3 s cadence time series. To display features on the fainter end, we averaged every 10 images (e.g., images 1–10, 11–20, 21–30, and so on). This results in a dataset with an effective cadence of 30 s. In this case, the time stamps are obtained from every sixth image from the original time series. For both these datasets, we then applied an unsharp-mask filter to sharpen the corresponding time-averaged images. All the images are displayed on a negative colour scale, so darker regions are brighter EUV emission features on the Sun, while lighter regions are fainter or darker regions in EUV.

We estimate the jet propagation speeds by computing the time difference in the intensity enhancements produced by a jet at two points in space along its length. We visually identify pixels along the path of a jet, that are separated in the plane-of-sky by more than two times the spatial resolution of HRI$_{EUV}$. Then we compute the mean intensity as a function of time, $I(t)$ around those points within a square region of ±1 to ±3 pixels from the unbinned data (i.e., original jitter-removed level-2 data with 3 s cadence). To identify intensity peaks at both locations due to the jet propagation, we fit a Gaussian model to the time series of mean intensities covering the period of the jet event. The Gaussian fitting used the MPFIT code, which implements the Levenberg-Marquardt least-squares minimization technique (*54*). We considered five fitting parameters (peak, centroid and standard deviation of the Gaussian, plus constant and linear terms). Uncertainties on intensity as a function of time, $\sigma_I(t)$, were included as input to the fitting procedure. We include the detector readout noise and the Poisson errors due to photon noise, but neglect the thermal noise. The uncertainties are computed as

$$\sigma_I(t) = \sqrt{r^2 + \frac{I(t) t_e \alpha}{n}}, \tag{S1}$$

where $r = 2$ DN is the detector readout noise, $t_\mathrm{e} = 1.65$ s is the exposure time, $\alpha = 6.85$ DN photon$^{-1}$ is the photons to DN conversion factor for the HRI$_\mathrm{EUV}$ detector (*55*), and $n$ is the sample size (i.e., number of pixels over which the intensity is averaged).

To determine the jet speed, $v$, we computed the separation between the centroids of the peaks of the two fitted Gaussians. This separation, $\Delta t$, is in units of time. The projected speed of the jet is then computed by $v = d/\Delta t$, where $d$ is the plane-of-sky distance between the two points along the jet. The standard error in $\Delta t$ is then propagated to estimate the standard error in $v$.

We measured cross-sectional widths of Y-shaped jets and plume jets from the frames from the 15 s effective cadence dataset (before unsharp masking). Intensities were interpolated onto coordinates defined by cross-cutting lines (shown in Fig.1C–L and Fig. 3). Then MPFIT was used to fit a Gaussian model to the intensity profile with five free parameters (as described above). We then define the cross-sectional width, $w$, as the full width at half maximum (FWHM) of the best-fitting Gaussian. Uncertainties in intensity are used to compute the standard error in $w$.

In Fig. 4 we followed a similar procedure, but used the frames from the 30 s effective cadence dataset. We defined a strip of 5 pixels wide and averaged the intensity across its width. Then we fitted a single Gaussian model with five free parameters to determine the FWHM of the jet feature as its cross-sectional width.

## Additional image processing

The 30 s effective cadence dataset was further processed in two ways to enhance the faint jet features. Firstly, we decomposed the images using the *à trous* wavelet transformation algorithm (*56*), starting with a triangular filter kernel, with elements [0.25, 0.5, 0.25]. For each image, we discard the highest frequency wavelet plane as noise, then sum the next two highest frequency planes to derive a sharpened, filtered version of the 30 s cadence time sequence. The resulting images are shown in Figs. S5A, S5C, S5E, S6A, S6C and S7A. These enhanced wavelet-filtered images show that the jet features propagate radially to greater distances from their source region, both in plume and inter-plume regions (see also Movie S10). Secondly, we derive a time sequence of running-difference images by subtracting every image from every second image in the 30 s effective cadence dataset. Snapshots from this running-difference image sequence are shown in Figs. S7B and S8 and Movie S11; the latter shows the predominantly southward propagation of the jet features.

A smooth-subtracted distance-time map (Fig. S9) was obtained by first interpolating HRI$_\mathrm{EUV}$ intensities from each snapshot from the 30 s effective cadence dataset, along a curved strip of pixels located above the limb (Fig. S4). Then we averaged the intensities across the strip and stacked the 1D array of intensities along the strip as a function of time. We smoothed this distance-time map using a running average in the spatial direction with a 50-pixel window. This smoothed image was then subtracted from the original image, resulting in the map shown in Fig. S9. This shows fine structure in the outflowing jets at the south pole over the plume and inter-plume regions.

## Calculation of kinetic energy flux

Here we estimate the jet energy content. The kinetic energy flux carried by a parcel of fluid moving with speed $v$ is

$$E_k = \frac{1}{2}\rho v^3, \tag{S2}$$

where $\rho$ is the fluid mass density. Based on our speed estimates, $v$ is ~100 km s$^{-1}$ (see main text).

We do not have direct spectroscopic density diagnostics to determine the jet plasma density, so instead estimate a lower limit. The ambient coronal hole density is about $2\times10^8$ cm$^{-3}$ (*57*). The Y-shaped jets, plume and inter-plume jets all appear 10% to 30% brighter than the surrounding coronal hole. The plasma temperature in coronal holes varies from about 0.8 MK in the plume regions to about 1.1 MK in inter-plume regions (*58*). The EUV jet features originating from these regions, visible in the HRI$_{EUV}$ passband which is sensitive to emission from 1 MK plasma, would then have correspondingly similar temperatures close to 1 MK. We therefore assume that the jet density is at least the ambient coronal hole density of $2\times10^8$ cm$^{-3}$. The kinetic energy flux content of a picoflare jet was then estimated from the values:

Electron density
$n_e = 2 \times 10^8$ cm$^{-3}$

Proton mass
$m_p = 1.67 \times 10^{-24}$ g

Fluid mass density
$\rho = m_p n_e = 3.34 \times 10^{-16}$ g cm$^{-3}$

Velocity
$v = 100$ km s$^{-1} = 10^7$ cm s$^{-1}$

Inserting $\rho$, and $v$ into equation S2 yields
$E_k = 1.67 \times 10^5$ g s$^{-3}$ [equivalent to erg cm$^{-2}$ s$^{-1}$]

This kinetic energy flux estimate corresponds to the plasma moving along the open magnetic field lines. To continually sustain the solar wind, there must be a continual supply of magnetic energy at the coronal base. Magnetic fields at the photosphere are recycled through emergence and cancellation of magnetic flux. Through this recycling, magnetic fields could channel energy to heat the upper atmosphere of the Sun. Previous estimates for the quiet Sun magnetic found a flux-loss rate of ~1150 maxwell (Mx) cm$^{-2}$ day$^{-1}$ at the photosphere (*59*), that would be sustained by emergence and cancellation of magnetic flux, due to magnetoconvection. Energy released during magnetic flux emergence and cancellation events, in principle, is sufficient to heat the entire quiet-Sun corona (*60, 61*). Surface magnetoconvection in coronal holes would be similar to that of the quiet Sun. Therefore, we expect similar flux emergence and cancellation rates also in the coronal hole regions. The magnetic energy liberated through flux emergence and cancellation events in coronal holes would then be sufficient to supply heated material to the solar wind (*62*).

Magnetic reconnection could also launch waves along the open fields in coronal holes (*63*), that might play a role in accelerating the solar wind. Irrespective of its generation mechanism, solar wind could adiabatically cool as it radially expands away from the Sun. It is thought that Alfvén waves which are broad-band in frequency, launched from coronal holes, could dissipate and maintain the ion temperature of the solar wind against adiabatic cooling (*64*). Thus picoflare jets launched by magnetic reconnection could be further sustained by Alfvén waves as they propagate away from the Sun.

**Contribution of picoflare jets to the solar wind**

We visually identified at least 120 picoflare jets in Fig. S8 and Movie S12, which we assume is a typical snapshot of the coronal dynamics. We take this to be the minimum number of jets in a coronal hole at any given time, then compute the lower limit of the filling factor of these jets in our observations. We estimate the on-disk coronal area (excluding brighter regions but including plumes) in our observations to be about 9000 $Mm^2$. Then the area covered by 120 picoflare jets, each with a width of 1 Mm (circular cross section) is about 94 $Mm^2$. This yields a filling factor of 0.01. This is about five times lower than the filling factor required for the observed picoflare jets to account for all the solar wind mass flux. Therefore, the picoflare jets we observe can account for at least 20% of the solar wind mass flux from coronal holes. Similarly, these picoflare jets can account for at least 20% of the energy flux to sustain the solar wind. We regard our estimate of filling factor of picoflare jets as a lower limit, because the picoflare jets we analyzed are the most common features in the observed coronal hole, compared to larger jets and eruptions (a large jet is visible in Fig. 1B and an eruption can be seen in Movie S1). A majority of such larger polar jets, often with X-rays signatures, most likely not escape from the Sun (*65*). Therefore, the remaining solar wind mass flux might not originate from these larger-scale features. Either even smaller scale jets, or even fainter picoflare jets, or the uniform component have to then supply the remaining mass flux. Our observations show continuous outflows from the inter-plume regions that cover a substantial portion of the coronal hole. Therefore, we argue that the observed picoflare jets and their fainter counterparts from inter-plume regions could be the dominant source of mass and energy to the solar wind.

## Supplementary Text

### Jet propagation

Figs. 2 and 4 show the intensity propagation in individual jet events by considering two spatial locations along the path of each jet. Here we further investigate the jet propagation by including an additional, third point along their path. Figs. S1 and S2 show Gaussian models fitted to the light curves from Figs 2 and 4, and to the average light curve obtained from a region between them. The three light curves from each of the individual jet events illustrate that the intensity variations propagate along these jets, from regions located closer to the jet-base to those farther away.

### Fainter veil-like emission from the observed coronal hole

To visualize the fainter emission from the observed coronal hole region, we used the 30 s effective cadence data (see Methods for details), and displayed the images on a log-scale to emphasize the regions with fainter emission in Fig. S3. Movies S3 and S4 reveal that this fainter veil-like emission displays predominantly southward propagating feature; a characteristic that makes it distinct from any static background structure.

### Y-shaped morphology and magnetic nature of the plume and inter-plume jets

Jets from plume and inter-plume regions shown in Figs. 3 and 4 do not have Y-shaped morphology at their base, which is unlike the jets shown in Fig. 1. Here we consider whether the lack of Y-shaped morphology at the base of plume and inter-plume jets could arise from line-of-sight or thermal effects. For this analysis we selected plume and inter-plume regions at the limb (designated as regions R4 to R8 in Fig S4).

Fig S5A shows two jet features emerging from the plume region R4 (Movie S5). Y-shaped morphology is visible at their bases. Compared to jet R4-1, the base of jet R4-2 has a lower intensity contrast with respect to the background.

Fig. S5C shows two jets with distinct morphological features from plume region R6. At its base, the smaller jet R6-1 exhibits a closed loop along with a bright dot, adjacent to which a jet emerges, with Y-shaped morphology. In contrast, the more elongated jet R6-2 appears to emerge at altitude over the limb, without a visible closed loop connected to the feature. However just below the lower tip of jet R6-2, there are multiple closed loops. Therefore, we expect a Y-shaped magnetic topology was present in the case of jet R6-2 as well, just not visible (with the same applies to other jets from the same region, shown in Movie S6). The absence of a Y-shaped morphology of jet R6-2 could be explained if the underlying closed loop connected to it was hotter or colder than ~1 MK (the peak of $HRI_{EUV}$ thermal response), so remained undetected by the instrument.

Fig. S5E shows two plume jets, both with Y-shaped morphology. The morphology of jet R8-2 is clearer than that of jet R8-1. In this case, because the base of each jet is closer to the solar surface, foreground and background coronal structures could obscure their visibility (Movie S7). There is a similar range of Y-shaped morphology visible for the inter-plume jets (Fig. S6; Movies S8–S9). Overall, we observe persistent jet activity, filling the plane-of-sky, in both the plume and inter-plume regions (Fig. S9).

Except for their orientation with respect to the observer, there is no reason to expect any fundamental physical difference between the on-disk plume region R2 (Fig. 3) and the limb plume regions R4, R6, and R8 (Fig. S5). In all these cases there are closed loop systems visible at the jet bases (in the case of R2, the closed loops are located just north of the region, see Fig. 1A, pointed by a red arrow in Movie S2). If interaction and interchange reconnection between closed and open magnetic field domains is the cause of the Y-shaped morphology in these jets, the apparent absence of that morphology in some of these jets could be related to line-of-sight or thermal effects. Therefore, although not directly observed, we expect that the outflows from plume R2 are also linked to an underlying Y-shaped magnetic topology. Similar arguments apply to the jets emerging from inter-plume regions R1-a, R1-b, R1-c, and R3-a.

We observed picoflare jets originating from a polar coronal hole that propagate outward nearly radially. But during periods of high solar activity, coronal holes often extend to lower latitudes and flank active regions. These picoflare jets could be common also in these low-latitude coronal holes adjacent to active regions. In such a case, channeling of picoflare jets from low-latitude coronal holes could be further governed by large-scale (global) magnetic topology of the Sun (*66*), so their outward propagation need not be radial.

Intensity propagations observed along plumes have been interpreted as either material flows (*19*) or wave motions (*67*). The same applies to intensity propagations detected along coronal fan loops that originate at the peripheries of active regions. At least in the case of coronal hole plumes, spectroscopic observations provide complementary evidence of interchange reconnection between open and closed magnetic fields at the base of coronal funnels that are thought to channel solar wind (*13*). Based on multi-thermal analysis of plumes, previous work has argued that the observed intensity variations along plumes are inconsistent with wave motions (*20*). In situ spacecraft observations in the heliosphere have been interpreted as evidence that interchange reconnection within coronal holes dominates driving of the solar wind (*68*). Our observations provide direct imaging evidence for the operation of small-scale interchange reconnection in the form of picoflare jets in coronal holes, which could supply mass and energy to the solar wind. The jet speeds we measured do not correspond to typical characteristic speeds associated with waves in coronal holes (see below). We therefore conclude that a wave explanation of the observed small-scale activity is unlikely.

## Characteristic speeds in coronal holes

The speed of sound in coronal gas of temperature, $T$, is given by

$$c_\text{s} = \sqrt{\frac{\gamma k T}{m}}, \tag{S3}$$

where $k = 1.38 \times 10^{-16}$ cm$^2$ g s$^{-2}$ K$^{-1}$ is the Boltzmann constant, $m = 1.67 \times 10^{-24}$ g is the mass of hydrogen, and $\gamma = 1.1$ is the effective adiabatic index in the solar corona (*69*). For plasma temperatures ranging from 0.8 MK in plumes to 1.1 MK in inter-plume regions, the speed of sound in coronal holes is in the range of ~85 km s$^{-1}$ to ~100 km s$^{-1}$.

Based on these sound speed values in coronal holes, the speeds of inter-plume jets in Fig. 4 are supersonic, so cannot be explained by slow magnetosonic waves. The Alfvén speed in coronal holes is ~1000 km s$^{-1}$ (*70*), much higher than our measured speeds. The fast magnetosonic speed is similar to the Alfvén speed. The jet speeds that we measured are slower than these faster wave modes, so we conclude they are not related.

## Observability of instabilities

Kelvin-Helmoholtz or resonant flow instabilities (*47*, *48*) induced by a velocity shear between neighboring high-speed picoflare jets could explain mid- to large-scale modulations of switchbacks observed in the solar wind. Vortex-like structures resembling Kelvin-Helmholtz instability, however, are only observed during unusual events, such as flare eruptions (*71*, *72*) and blowout jets (*73*), which are much larger than the picoflare jets that we observed. Numerical models of coronal loops predict that Kelvin-Helmholtz instability induced by shear flows might be common and that it operates on scales that are smaller than the observations were performed, located at the edges of coronal loops (*74*). Any twist in the magnetic fields would make any Kelvin-Helmholtz instability in the solar corona harder to detect (*75*). Inner heliospheric observations of scales of switchbacks provide complementary information about the scales of instabilities at the origin of solar wind in the low corona. Previous work has inferred that the processes leading to a mid-scale modulation of a patch of switchbacks are operating on granular scales on the Sun (*29*). This means that an individual switchback, within a given patch of mid-scale modulation in the solar wind, itself might originate from even smaller spatial scales on the Sun. We therefore do not expect to spatially resolve the source of an individual switchback in the corona with our observations.

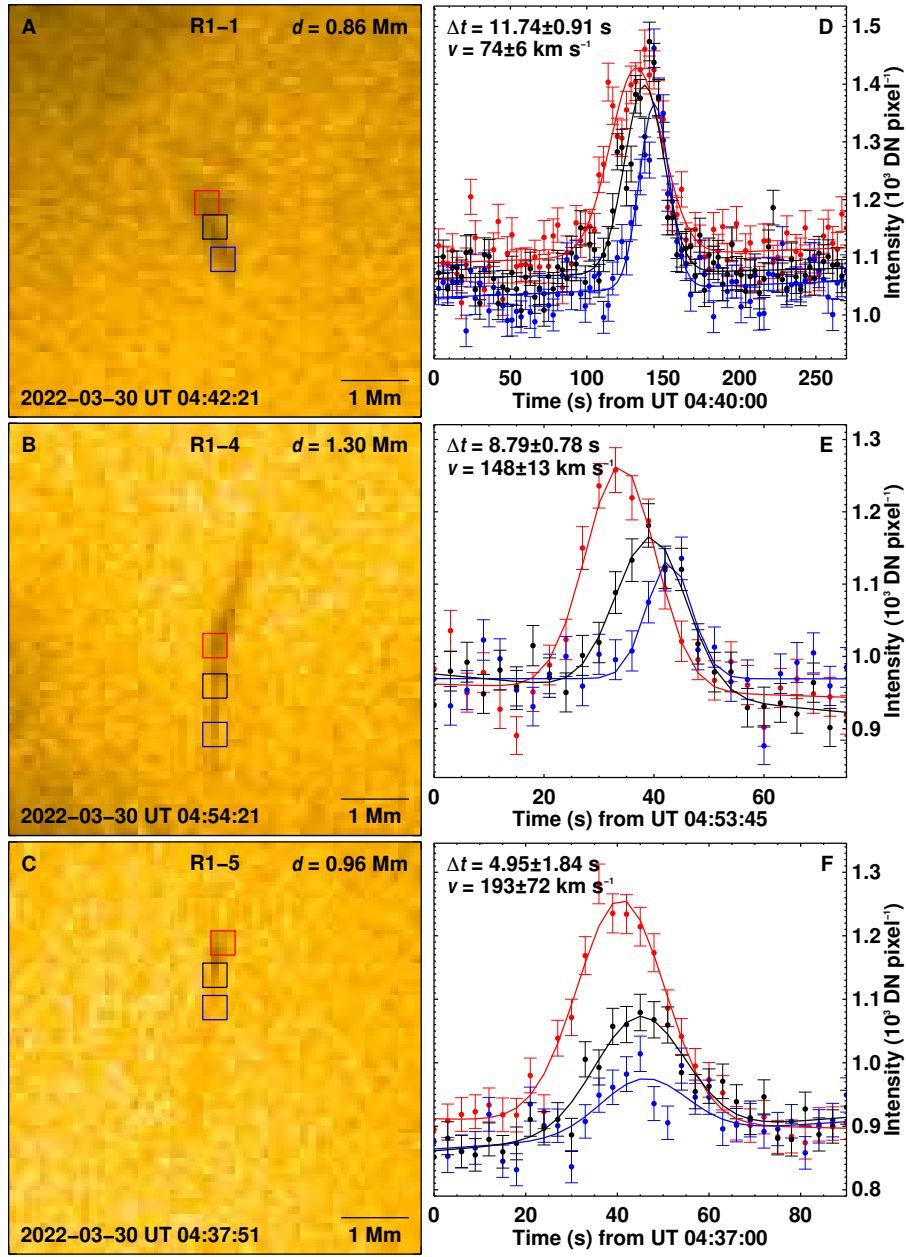

**Fig. S1. Propagating intensity disturbances along Y-shaped jets.** Same as Fig. 2, but with an additional location (black square) along the jet path between the red and blue squares. The parameters, $d$, $\Delta t$, and $v$ are the same as in Fig. 2. See text for details.

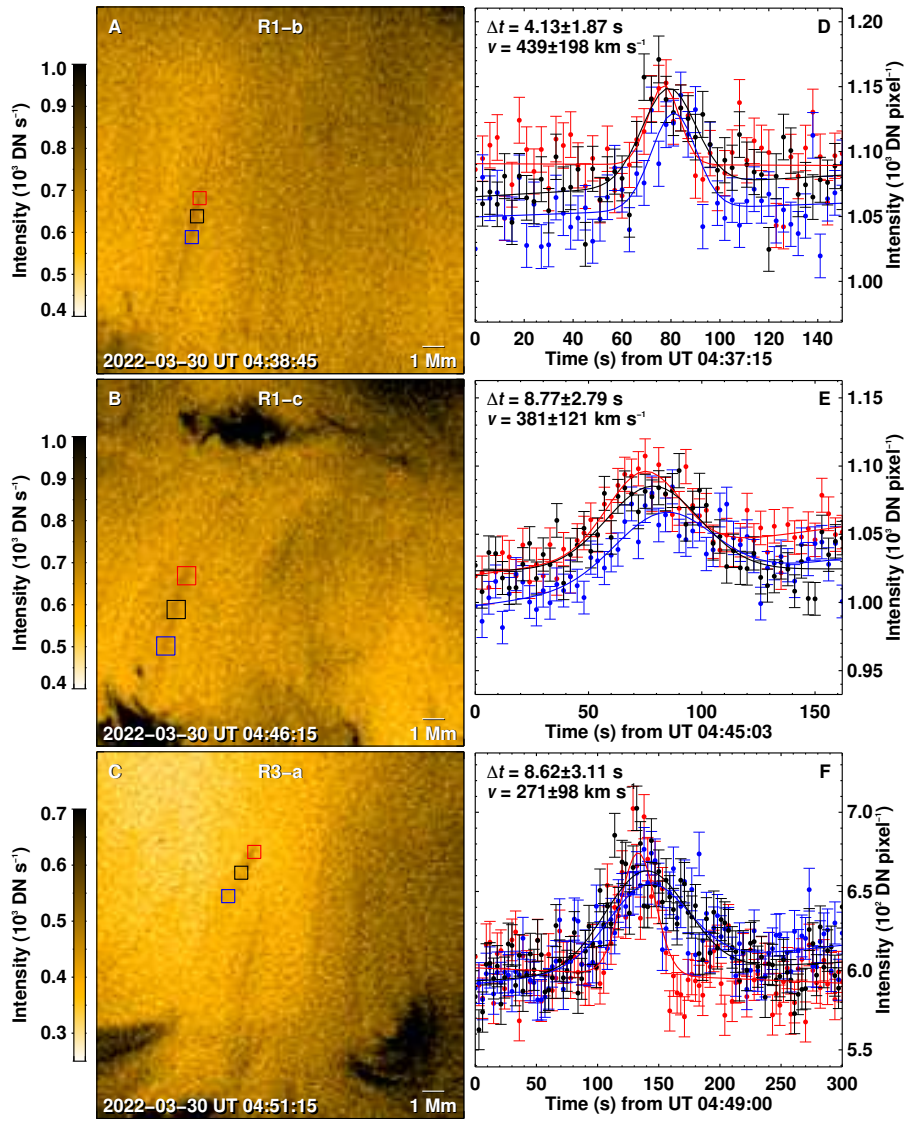

**Fig. S2. Propagating intensity disturbances along faint inter-plume jets.** Same as Fig. S1, but for the events in Fig. 4.

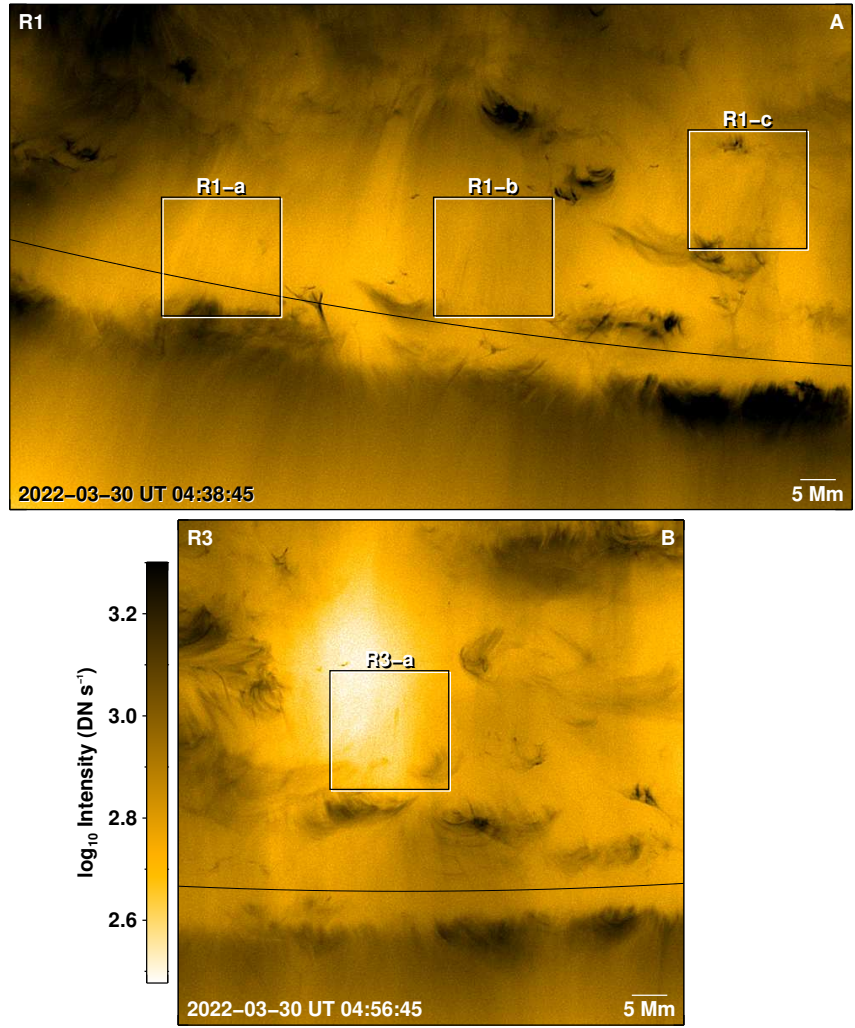

**Fig. S3. Fainter veil-like emission from the coronal hole.** Panels A and B show regions R1 and R3 from Fig. 1A. In these negative images, intensities are logarithmically scaled to the same brightness scale. The smaller sub-regions (R1-a through R1-c and R3-a) outline areas from which some of this fainter emission originates in the form of transient jets. The snapshots are from the 30 s effective cadence data (see text for details). The solar limb in both the panels is identified with a solid curve. Animated versions of panels A and B are shown in Movies S3 and S4 respectively.

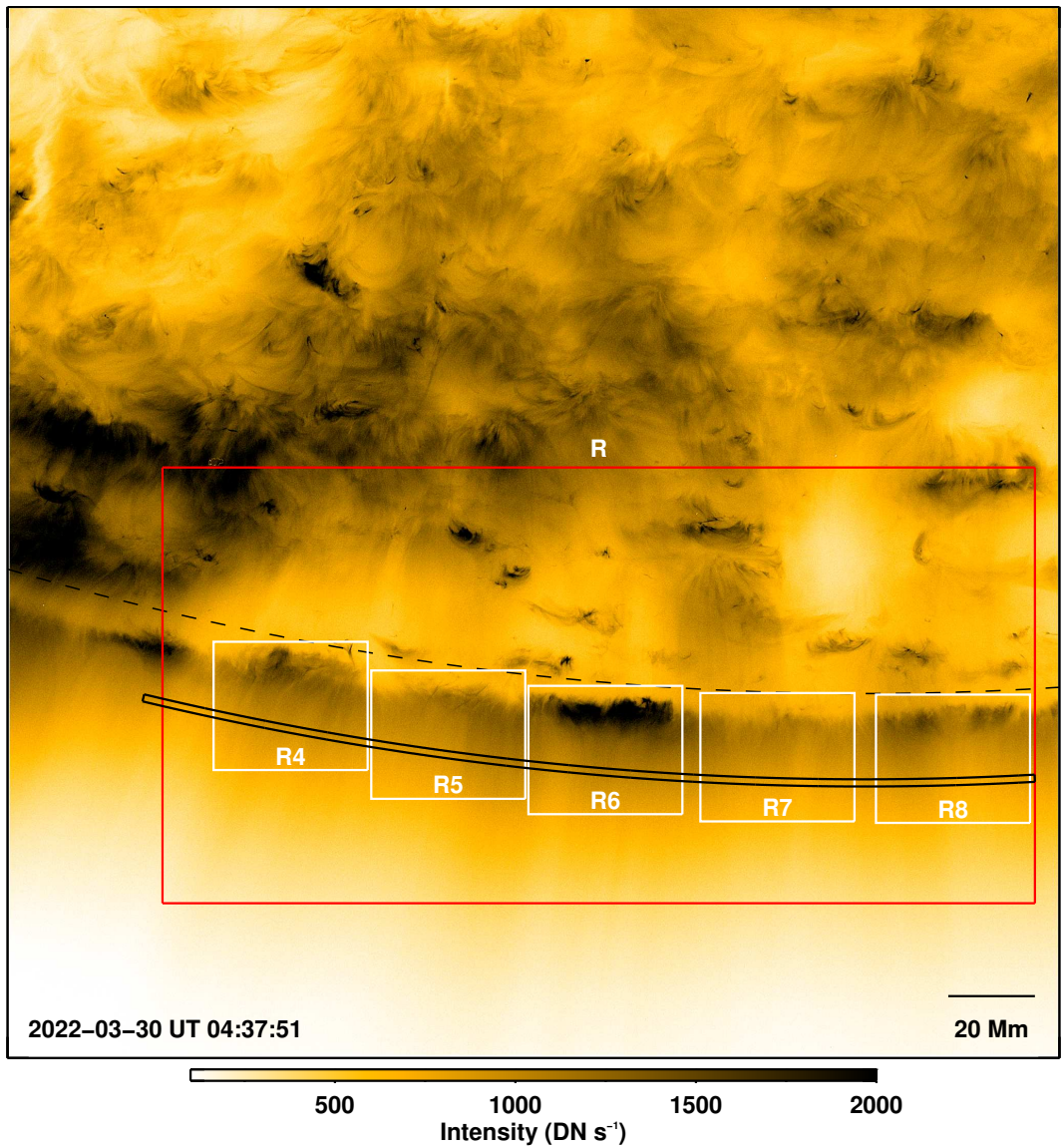

**Fig. S4. Additional examples of plumes and inter-plumes and the coronal hole context.** Same image as Fig. 1A but with regions overlain. Boxes R4, R6, and R8 cover plume regions at the southern limb (Fig. S5). The interspersed boxes R5 and R7 cover inter-plume regions (Fig. S6). The larger red box R outlines the coronal hole (Fig. S7). The black strip, ~20 Mm above the solar disk, parallel to the limb, is used to create the distance-time plot shown in Fig. S9.

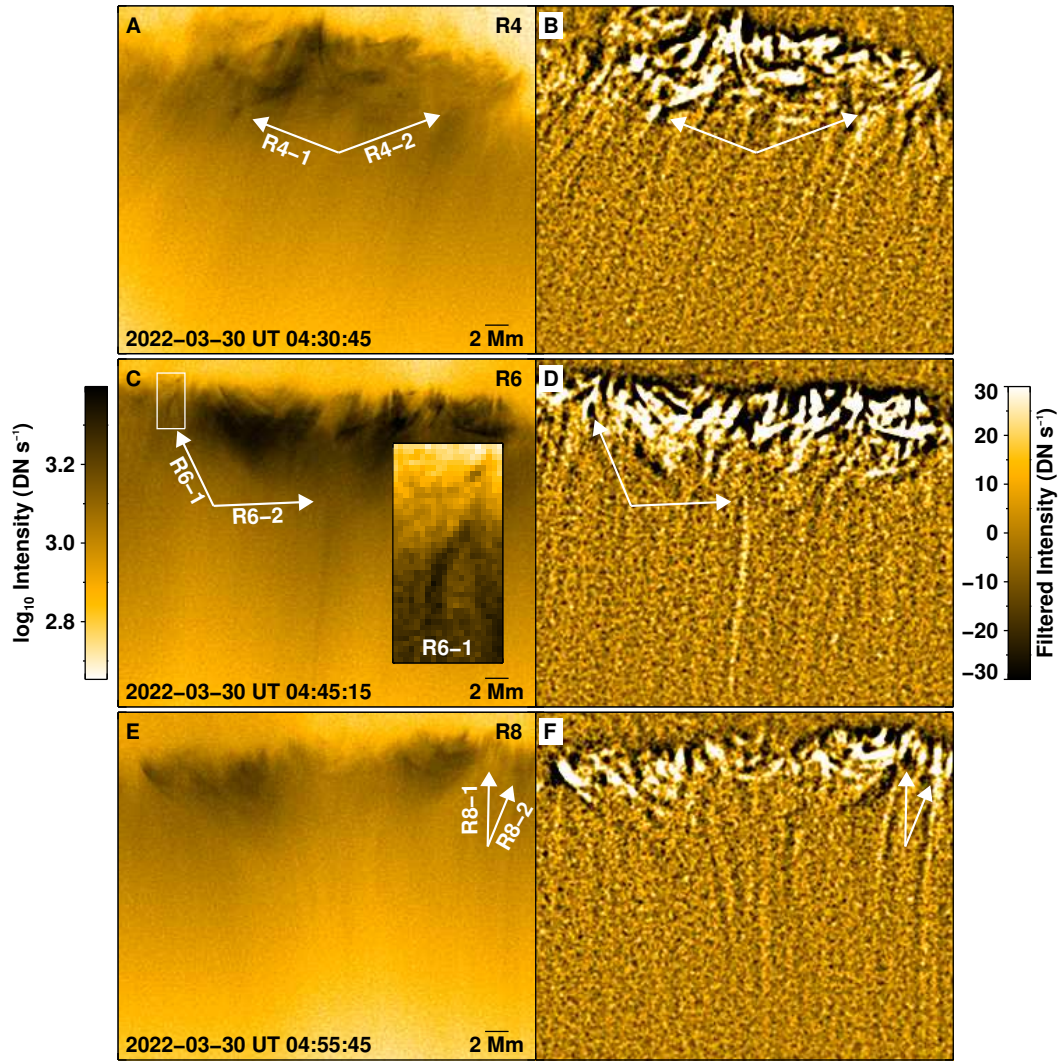

**Fig. S5. Jet from plumes.** Panels **A**, **C**, and **E** cover plume regions R4, R6, and R8, respectively, marked in Fig. S4. In each panel, arrows point to jets originating from the corresponding plume region. Intensities in panels **A**, **C**, and **E** are logarithmically scaled to the same brightness scale. The inset in panel C, with intensity on square-root-scale, is a zoom into the region outlined by a white rectangle in that panel (showing jet R6-1). The snapshots in panels **B**, **D**, and **F** are enhanced wavelet-filtered images (see text) showing the fine structuring of the jets. These panels are on a non-inverted color scale, so regions that appear bright in the image are also bright in EUV on the Sun. Animated versions of panels **A**-**B**, **C**-**D**, and **E**-**F** are shown in Movies S5, S6, and S7 respectively.

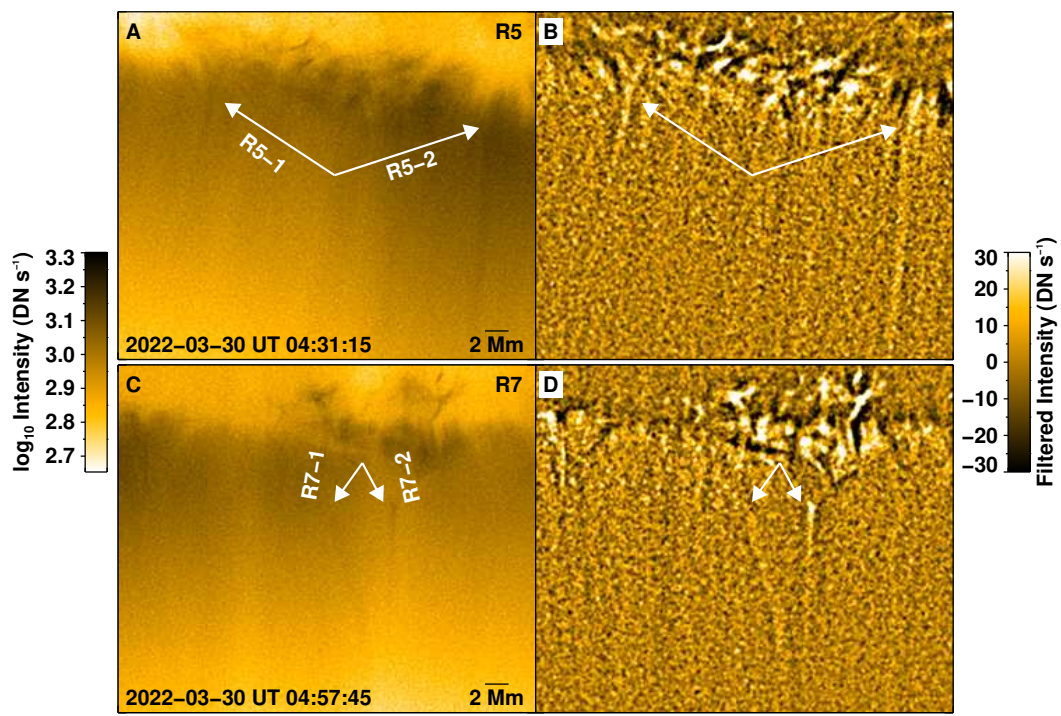

**Fig. S6. Jet from inter-plumes.** Same as Fig. S5, but for the inter-plume regions R5 and R7 marked in Fig. S4. Animated versions of panels **A**-**B** and **C**-**D** are shown in Movies S8 and S9.

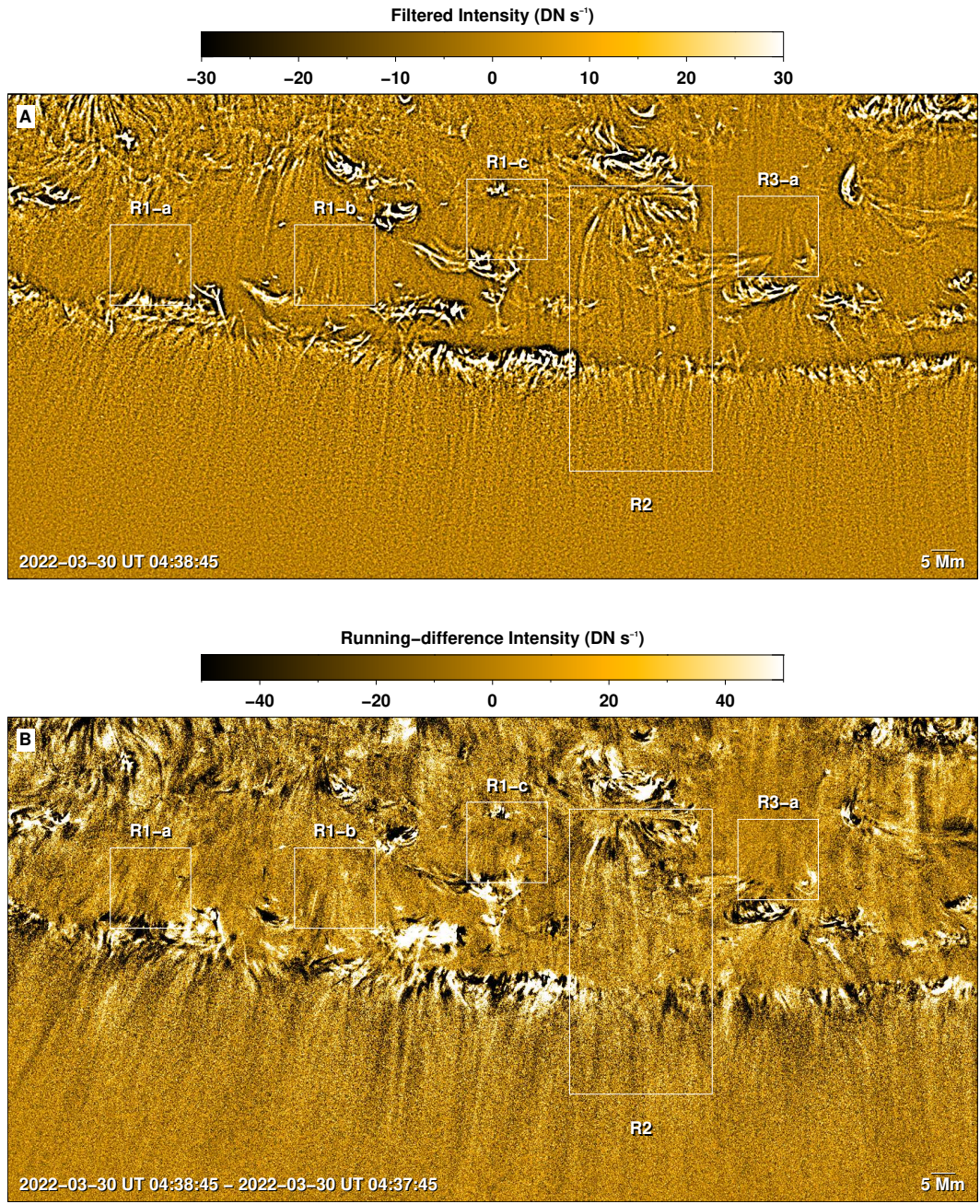

**Fig. S7. Coronal hole context.** (**A**) An enhanced HRI$_{EUV}$ image (see text) of region R (Fig. S4) showing finely-structured elongated coronal features at the south pole This image is on a non-inverted color scale, so regions that appear bright in the image are also bright in EUV on the Sun. (**B**) A running-difference image (see text) highlighting the flows in the coronal hole. Boxes R1-a, R1-b, R1-c, and R3-a cover inter-plume regions, whereas box R2 covers the plume region (see also Fig. 1). Animated versions of panels **A** and **B** are shown in Movies S10 and S11.

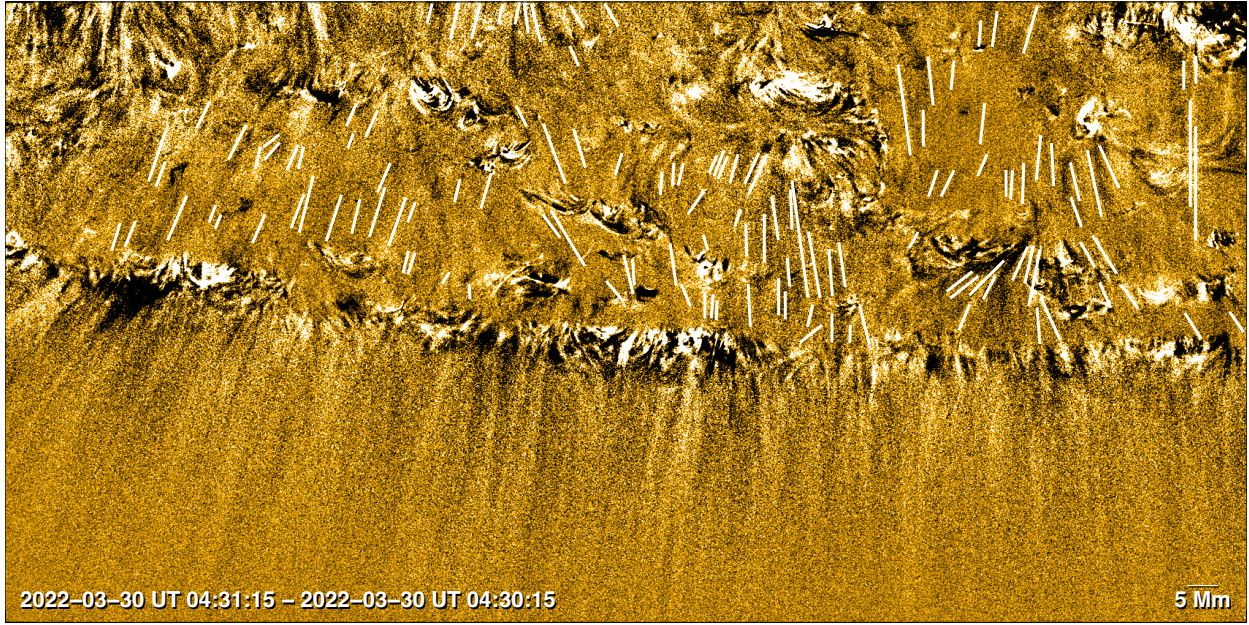

**Fig. S8. Widespread jets from the coronal hole.** Running difference image (see text) from the beginning of the observed time series. White lines mark the locations of 126 visually identified jets in this frame. An animated version of this figure but alternating with and without markings of visually identified jets is shown in Movie S12.

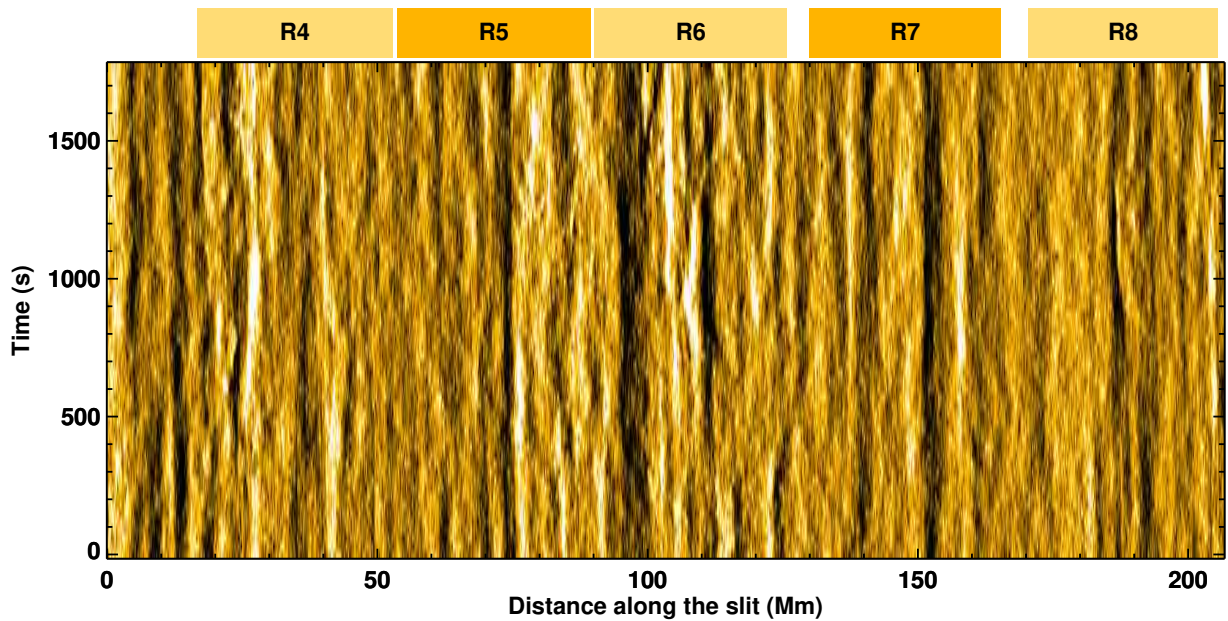

**Fig. S9. Persistent and widespread outflows from the coronal hole.** The smooth-subtracted distance-time map (see text) of outflowing jet features, crossing the strip above the limb marked in Fig. S4. Colored rectangles at the top indicate the extent of the plume and inter-plume regions shown in Fig. S4.

**Table S1. Width of Y-shaped jets.** Width, $w$, of Y-shaped jets shown in Fig. 1.

| Fig. 1 (panel) | Jet width, $w$ (km) |
|---|---|
| C | 307±42 |
| D | 245±26 |
| E | 303±45 |
| F | 291±36 |
| G | 229±28 |
| H | 331±39 |
| I | 373±80 |
| J | 283±36 |
| K | 232±41 |
| L | 200±46 |

**Table S2. Parameters of Y-shaped jets.** Locations along each jet, separated by a plane-of-sky distance, $d$, the separation between the centroids of the two Gaussian peaks, $\Delta t$, and the speed, $v = d/\Delta t$ with their standard errors shown in Fig. 2.

| Fig. 2 (jet) | Separation, $d$ (Mm) | Time difference, $\Delta t$ (s) | Speed, $v$ (km s$^{-1}$) |
|---|---|---|---|
| R1-1 | 0.86 | 11.74±0.91 | 74±6 |
| R1-4 | 1.30 | 8.79±0.78 | 148±13 |
| R1-5 | 0.96 | 4.95±1.84 | 193±72 |

**Table S3. Width of plume-related jets.** Width, $w$, of plume-related jets shown in Fig. 3.

| Fig. 3 (panel) | Jet width, $w$ (km) |
|---|---|
| A | 460±79 |
| B | 556±137 |
| C | 457±76 |
| D | 595±74 |
| E | 305±51 |
| F | 358±50 |
| G | 266±45 |
| H | 227±56 |

**Table S4. Parameters of inter-plume jets.** Width, $w$, locations along each jet, separated by a plane-of-sky distance, $d$, the separation between the centroids of the two Gaussian peaks, $\Delta t$, and the speed, $v = d/\Delta t$ with their standard errors shown in Fig. 4.

| Fig. 4 (region) | Jet width, $w$ (km) | Separation, $d$ (Mm) | Time difference, $\Delta t$ (s) | Speed, $v$ (km s$^{-1}$) |
|---|---|---|---|---|
| R1-b | 243±54 | 1.81 | 4.13±1.87 | 439±198 |
| R1-c | 417±100 | 3.34 | 8.77±2.79 | 381±121 |
| R3-a | 197±53 | 2.34 | 8.62±3.11 | 271±98 |

## Captions for Movies S1 to S12

**Movie S1:** Animated version of Fig. 1B, showing widespread Y-shaped jets in the observed coronal hole.

**Movie S2:** Animated version of Fig. 3 showing prevalence of plume jets from region R2. The red arrow points to the base of the plume.

**Movie S3:** Animated version of Fig. S3A, showing the ubiquitous faint veil-like emission and faint jets from inter-plume regions in R1.

**Movie S4:** Animated version of Fig. S3B.

**Movie S5:** Animated version of Fig. S5A-B, showing Y-shaped jets observed over plume region R4 at the limb.

**Movie S6:** Animated version of Fig. S5C-D.

**Movie S7:** Animated version of Fig. S5E-F.

**Movie S8:** Animated version of Fig. S6A-B.

**Movie S9:** Animated version of Fig. S6C-D.

**Movie S10:** Animated version of Fig. S7A, showing finely-structured elongated jet features emerging throughout the coronal hole.

**Movie S11:** Animated version of the running difference Fig. S7B, showing outward propagating jet features from the coronal hole.

**Movie S12:** Same as Fig. S8, but alternating with and without markings of visually identified jets (white lines).